\documentclass[aps, prl, twocolumn, preprintnumbers, superscriptaddress, nofootinbib, floatfix]{revtex4-1}

\usepackage{amsmath}
\usepackage{bm}
\usepackage{amsfonts}
\usepackage{graphicx}
\usepackage{hyperref}
\usepackage{color}
\usepackage{epsfig}

\hypersetup{
colorlinks=true,
linkcolor=blue,
linktoc=page,
citecolor=blue,
urlcolor=blue}

\begin{document}

\title{\bf Cyclic Cosmology and Geodesic Completeness}

\newcommand{\FIRSTAFF}{\affiliation{Department of Physics, University at Buffalo, Buffalo, NY 14260, USA}}

\author{William H. Kinney}
\email[Electronic address: ]{whkinney@buffalo.edu}
\FIRSTAFF
\author{Nina K. Stein}
\email[Electronic address: ]{ninastei@buffalo.edu}
\FIRSTAFF

\date{November 11, 2021}
\begin{abstract}
We consider recently proposed bouncing cosmological models for which the Hubble parameter is periodic in time, but the scale factor grows from one cycle to the next as a mechanism for shedding entropy. Since the scale factor for a flat universe is equivalent to an overall conformal factor, it has been argued that this growth corresponds to a physically irrelevant rescaling, and such bouncing universes can be made perfectly cyclic, extending infinitely into the past and future. We show that any bouncing universe which uses growth of the scale factor to dissipate entropy must necessarily be geodesically past-incomplete, and therefore cannot be truly cyclic in time. 
\end{abstract}

\maketitle

\section{Introduction}

A widely discussed alternative to cosmological inflation is the idea of a bouncing cosmology, in which the current epoch of expansion is preceded by a period of contraction, during which the universe is dynamically driven to a state of flatness and homogeneity. The possibility of a truly cyclic cosmology, which periodically oscillates between expansion and contraction, is especially appealing, since it removes the need for a cosmological boundary condition as in the case of the big bang singularity, or inflation, which was shown by Borde, Guth, and Vilenkin (BGV) to be geodesically past incomplete \cite{Borde:2001nh}. It has been known since work by Tolman in the 1930s that cyclic cosmology suffers from the problem of entropy growth \cite{Tolman:1931fei}. The second law of thermodynamics operates identically whether the universe is expanding or contracting, and the growth of entropy with time prevents cyclic behavior. Black hole formation is a particularly difficult problem: black holes formed during one period of expansion and contraction inevitably lead to a strongly inhomogeneous bounce, regardless of the background dynamics. A model recently proposed by Ijjas and Steinhardt (IS) \cite{Ijjas:2019pyf} suggests a solution to the entropy problem in cyclic cosmology \cite{Ijjas:2021zwv}. In this model, the Hubble parameter for the expansion rate is periodic in time (Fig. 1), but the cosmic scale factor grows exponentially from one oscillation to the next.  The expanding phase of the universe at late times is exponential, which exponentially dilutes the entropy within the cosmological horizon. During the subsequent contracting phase, the equation of state of the universe is extremely stiff, with $p \gg \rho$, resulting in slow contraction, with a Hubble parameter near zero and a slowly varying scale factor. This means that the entropy redshifted outside the cosmological horizon during the phase of accelerated expansion never falls back \textit{into} the horizon, and is argued to be a solution to the entropy problem in cyclic cosmology, since the scale factor in a flat Friedmann-Robertson-Walker spacetime is an irrelevant overall conformal factor, and can be arbitrarily rescaled. 

In this paper, we examine the IS cyclic model from the standpoint of geodesic completeness. While the Hubble parameter may in principle be discontinuous across a bounce connecting a contracting phase to an expanding one, the scale factor will in general be continuous. This means that spacetime geodesics can be traced continuously through the bounce solution. The IS model postulates a cosmology with alternating expanding and contracting phases with period $T$, such that the Hubble parameter is exactly periodic,
\begin{equation}
    H\left(t + T\right) = H\left(t\right). 
\end{equation}
The scale factor, however, grows exponentially from one cycle to the next, as a mechanism for shedding the entropy produced in each cycle,
\begin{equation}
    a\left(t + T\right) = e^N a\left(t\right),
\end{equation}
where $N$ is a constant chosen to be large enough to ensure a sufficiently smooth boundary condition at the bounce.  Calling $H_+\left(t\right) > 0$ the Hubble parameter during expansion, and $H_-\left(t\right) < 0$ the corresponding rate of contraction, the condition for entropy dissipation is 
\begin{equation}
    \int_0^\tau{H_+\left(t\right) dt} \gg - \int_\tau^T{H_-\left(t\right) d t},
    \label{eq:EntropyCondition}
\end{equation}
where $\tau$ is the duration of the expanding phase. Despite periodic $H\left(t\right)$, a difference in average equation of state between expanding and contracting phases results in growth of the scale factor. Taking the continuity equation for the cosmic density $\rho\left(t\right)$,
\begin{equation}
    \dot\rho + 3 H \left(1 + w\right) \rho = 0,
\end{equation}
where the Hubble parameter is given by
\begin{equation}
    H^2 = \frac{1}{3 M_\mathrm{P}^2} \rho,
\end{equation}
and the equation of state is 
\begin{equation}
    p = w \rho,
\end{equation}
we can define the expanding phase by an average equation of state $w_+ \simeq -1$, and the contracting phase by $w_- \gg 1$. Then $\dot\rho_+ \ll H \rho_+$, and
\begin{equation}
    H_+^2 = \frac{1}{3 M_\mathrm{P}^2} \rho_+ \simeq \mathrm{const.}
\end{equation}
Similarly, $\dot\rho_- \gg H \rho_-$, and
\begin{equation}
    H_-^2 = \frac{1}{3 M_\mathrm{P}^2} \rho_- \ll H_+^2. 
\end{equation}
A simplified example is an expanding phase dominated by a cosmological constant $w_+ = -1$, so that $H_+ = \mathrm{const.}$, and $a_+\left(t\right) = e^{H t}$. The contracting phase is dominated by a scalar field $\phi$ with negative potential, so that
\begin{equation}
    \rho_- = \frac{1}{2} \dot\phi^2 + V\left(\phi\right) \simeq 0.
\end{equation}
Approximately vanishing density means that $H_- \simeq 0$, with the result that $a_-\left(t\right) \simeq \mathrm{const.}$ This is the well-studied case of ``ekpyrotic'' contraction (for a review, see  Ref. \cite{Lehners:2008vx}.) In this limit,
\begin{equation}
    N_+ = \int_0^\tau{H_+\left(t\right) dt} \simeq H_+ \tau,
\end{equation}
and
\begin{equation}
    N_- = \int_\tau^T{H_-\left(t\right) dt} \simeq 0,
\end{equation}
which trivially satisfies the general condition (\ref{eq:EntropyCondition}). In this limit, entropy is diluted during a period of late exponential expansion (inflation), with a subsequent slow contraction phase during which the scale factor remains approximately constant. 

At the bounce itself, the Hubble parameter undergoes a sudden transition from small and negative to large and positive, which necessarily violates the Null Energy Condition, requiring modification to General Relativity. In the limit that the Hubble parameter undergoes a discontinuity at the bounce, the scale factor is nonetheless well-behaved and continuous. This means that geodesic spacetime paths can be traced through multiple cycles of expansion, contraction, and bounce. A necessary condition for infinite cyclic behavior is that the corresponding spacetime be \textit{geodesically complete}, defined as having the property that all geodesic world lines through spacetime be both future- and past-infinite as measured by proper time along the world line from any finite time $t$,
\begin{equation}
    \int_{-\infty}^t{d s} \rightarrow \infty, \qquad \int_t^{+\infty}{ds} \rightarrow \infty. 
\end{equation}
 Because the scale factor varies exponentially over the number of cycles, the spacetime is nonsingular over the range $t = [-\infty,\infty]$, with
 \begin{equation}
     a\left(t \rightarrow -\infty\right) \propto e^{N} \rightarrow 0
 \end{equation}
 as $N \rightarrow -\infty$. The spacetime is trivially both future- and past-infinite for comoving observers. However, geodesic completeness means that \textit{all} geodesic world lines must be future- and past-infinite, including those of non-comoving observers. Inflationary spacetimes were shown to be generically geodesically \textit{incomplete} by BGV in Ref. \cite{Borde:2001nh}. In the next section, we apply the BGV theorem to the IS model, and show that it, like the inflationary spacetime, is geodesically incomplete, and for the same reason.

\section{Geodesic completeness}
\label{sec:Completeness}

In this section, we show that any bouncing spacetime which satisfies the condition (\ref{eq:EntropyCondition}) for entropy dissipation must be geodesically past-incomplete. To do so, consider a geodesic test particle in a Friedmann-Robertson-Walker spacetime, with metric
\begin{equation}
    ds^2 = dt^2 - a^2\left(t\right) d \mathbf{x}^2.
\end{equation}
The Hubble parameter is defined as
\begin{equation}
    H \equiv \pm \frac{\dot a}{a},
\end{equation}
with the $\left(+\right)$ branch corresponding to an expanding phase, and the $\left(-\right)$ branch corresponding to a contracting phase. Taking a metric $g_{\mu \nu} = \mathrm{diag.}\left(1,-a^2,-a^2,-a^2\right)$, timelike normalization of the four-velocity $u^\mu u_\mu = 1$ gives
\begin{equation}
    u^\mu u_\mu = g_{\mu \nu} \frac{d x^\mu}{ds} \frac{d x^\nu}{ds} = \left(\frac{d t}{ds}\right)^2 - a^{2}\left(t\right) \left\vert \frac{d \mathbf{x}}{d s}\right\vert^2 = 1.
\end{equation}
The geodesic equation for the motion of a test particle is
\begin{equation}
    \frac{d}{d s}\left[a^2\left(t\right) \frac{d \mathbf{x}}{d s}\right] = 0,
\end{equation}
so that we can define an integration constant $v_0$ such that \cite{Aguirre:2001ks}
\begin{equation}
a^2\left(t\right) \left\vert \frac{d \mathbf{x}}{d s} \right\vert \equiv v_0 = \mathrm{const.}
\end{equation}
We then have
\begin{equation}
    \left(\frac{d t}{ds}\right)^2 = 1 + v_0^2 a^{-2}\left(t\right). 
\end{equation}
The differential proper time $ds$ along the geodesic of the test particle can then be written in terms of the coordinate time $dt$ as 
\begin{equation}
    ds = \frac{dt}{\sqrt{1 + v_0^2 a^{-2}\left(t\right)}}. 
\end{equation}

Consider the simple case of de Sitter expansion,
\begin{equation}
    H = \mathrm{const.} \Rightarrow a\left(t\right) = e^{H t}.
\end{equation}
Although the scale factor is well-behaved at any past finite time, vanishing only in the limit $t \rightarrow -\infty$, it is straightforward to show that the spacetime is geodesically incomplete by integrating the proper time backward from time $t=0$ to $t = -\infty$ \cite{Aguirre:2001ks},
\begin{eqnarray}
    \Delta s &&= \int_{-\infty}^{0}{\frac{dt}{\sqrt{1 + v_0^2 e^{-2 H t}}}}\cr
    &&= \frac{1}{2 H} \int_0^{\infty}{\frac{du}{\sqrt{1 + v_0^2 e^{u}}}},\qquad u \equiv - 2 H t\cr
    &&= \frac{1}{H} \tanh^{-1}{\left(\sqrt{1 + v_0^2 e^{-2 H t}}\right)}\Bigg\vert_{t \rightarrow -\infty}^{t = 0}\cr
    &&= \frac{1}{2 H} \ln \left(\frac{\sqrt{1 + v_0^2} + 1}{\sqrt{1 + v_0^2} - 1}\right).
    \label{eq:deSitter}
\end{eqnarray}
Since this integral is finite for any nonzero $v_0$, the past proper time along the geodesic is finite, and the space is therefore geodesically incomplete. This is a simple special case of the BGV theorem, which applies in general to any inflationary spacetime. 

\begin{figure}
    \centering
    \includegraphics[width=0.9 \columnwidth]{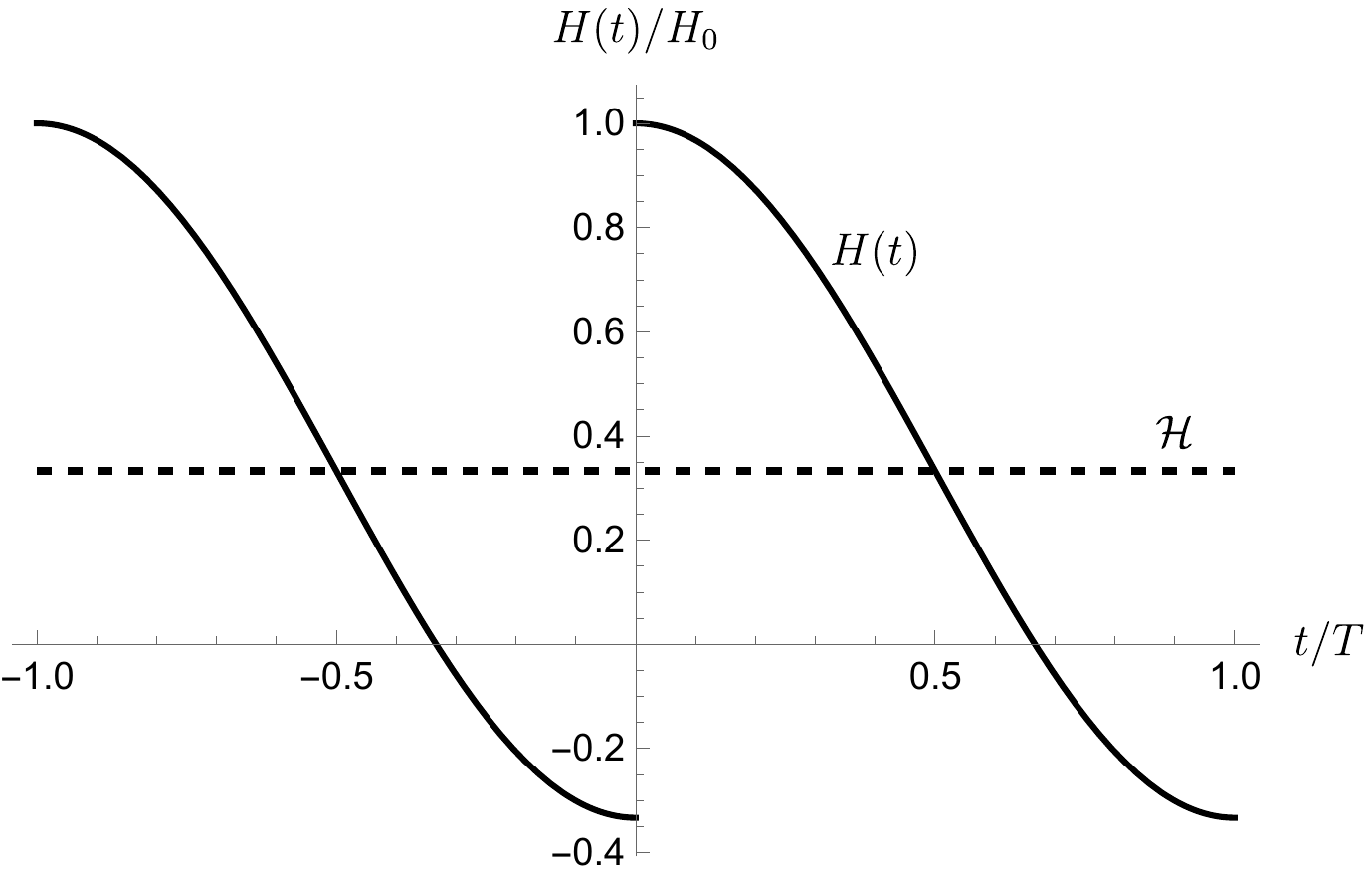}
    \caption{A schematic sketch of $H\left(t\right)$ for the Ijjas/Steinhardt model of cyclic cosmology \cite{Ijjas:2019pyf,Ijjas:2021zwv}. The dashed line shows the mean expansion rate $\mathcal{H}$.}
    \label{fig:Hvst}
\end{figure}

\begin{figure}
    \centering
    \includegraphics[width=0.9 \columnwidth]{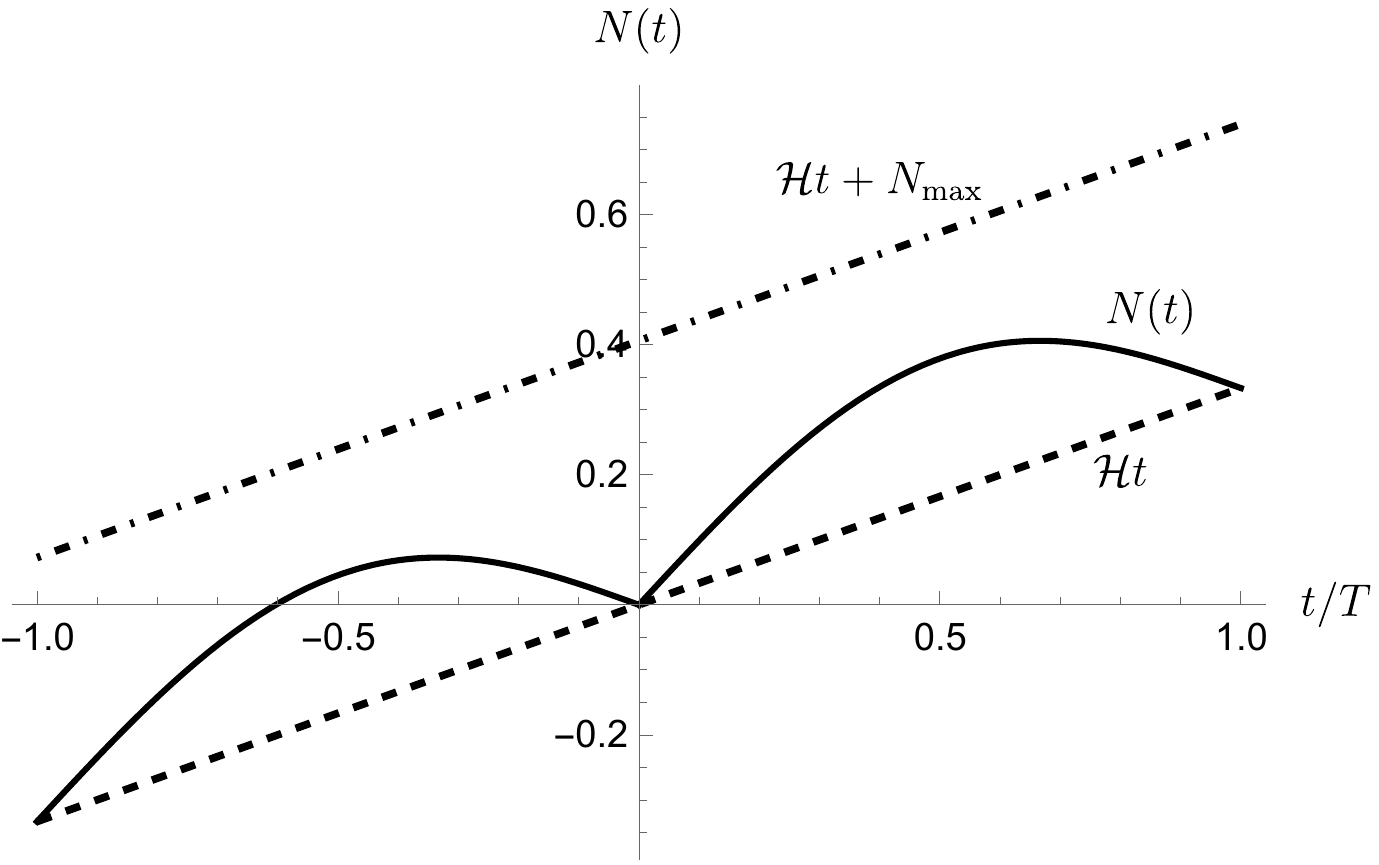}
    \caption{The number of e-folds $N\left(t\right) = \int{H\left(t\right) d t}$ vs time for the cyclic case (solid line), and the bounding de Sitter solutions (dashed,dotted).}
    \label{fig:Nvst}
\end{figure}

\begin{figure}
    \centering
    \includegraphics[width=0.9 \columnwidth]{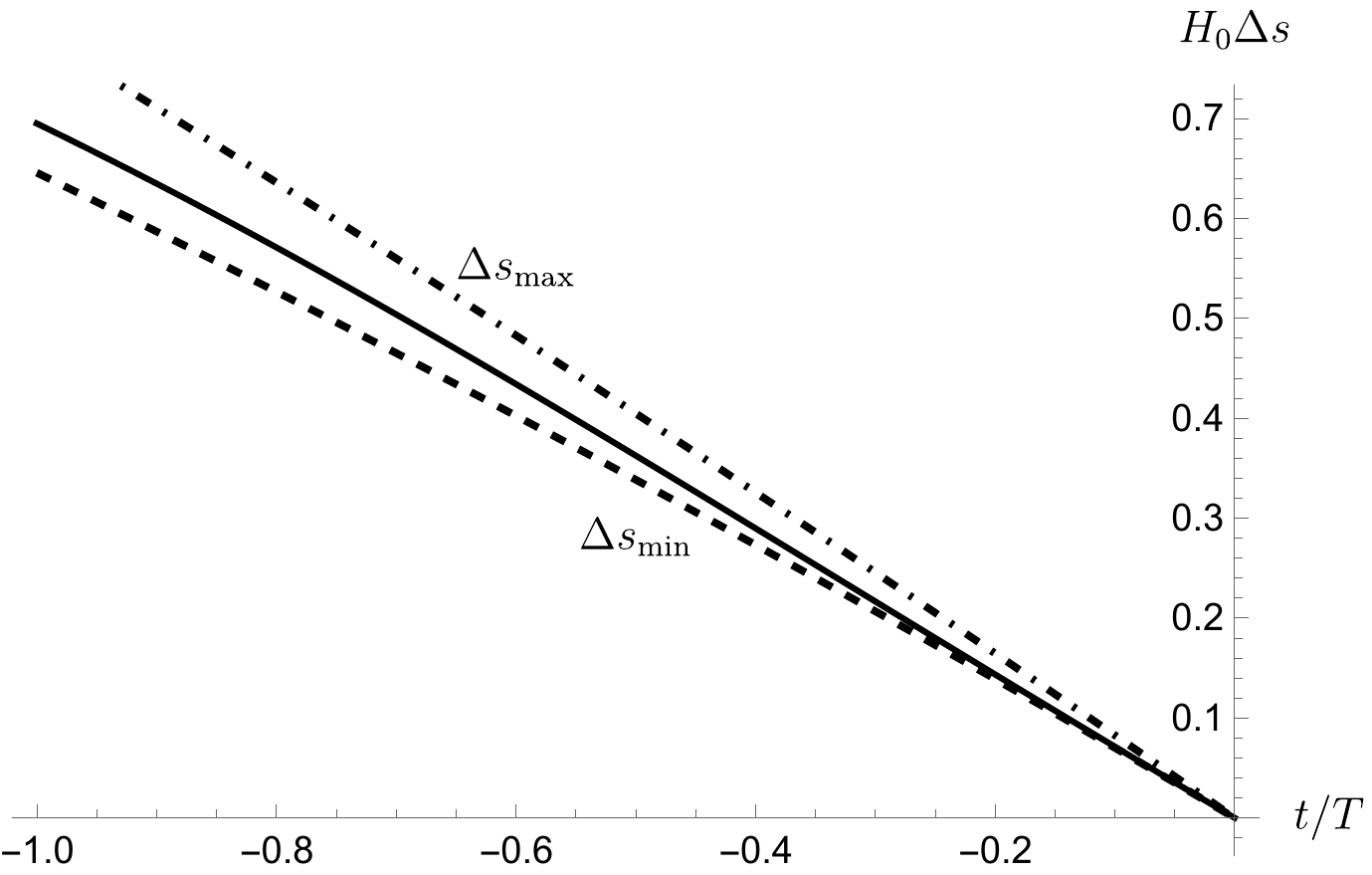}
    \caption{$\Delta s$ for the cyclic case (solid line), and the bounding de Sitter solutions $\Delta s_{\mathrm{min}}$ (dashed), $\Delta s_{\mathrm{max}}$ (dotted).}
    \label{fig:svst}
\end{figure}

Now consider a universe with $H(t)$ periodic with period $T$, which satisfies the condition (\ref{eq:EntropyCondition}) for entropy dissipation, $N_+ + N_- > 0$. We assume that the bounce is rapid enough that we can take $H\left(t\right)$ to be discontinuous across the bounce at the boundary of the interval, with scale factor $a\left(t\right)$ continuous, as shown in Fig. \ref{fig:Hvst}.  The condition for entropy dissipation (\ref{eq:EntropyCondition}) means there must exist a mean expansion rate $\mathcal{H}$ on the interval $0 < t < T$ such that
\begin{equation}
    \mathcal{H} \equiv \frac{1}{T} \int_{0}^T{H\left(t'\right) dt'} > 0. 
\end{equation}
Requiring that the Null Energy Condition holds everywhere except the bounce is equivalent to the condition that $H\left(t\right)$ be constant or decreasing,
\begin{equation}
    \frac{d H\left(t\right)}{d t} \leq 0,
    \label{eq:NEC}
\end{equation}
for $t$ in the interval $0 < t < T$. Then
\begin{equation}
   0 < \mathcal{H} t < \int_{0}^t{H\left(t'\right) dt'}
\end{equation}
for all $0 < t < T$. This means that
\begin{equation}
    e^{\mathcal{H} t} < \exp\left[\int_{0}^t{H\left(t'\right) dt'}\right]
\end{equation}
for all $0 < t < T$. The Null Energy Condition (\ref{eq:NEC}) likewise requires that there exist a finite $N_{\mathrm{max}}$
\begin{equation}
    N_{\mathrm{max}} \equiv \mathrm{Max}\left[\int_{0}^t{H\left(t'\right) dt'}\right],
\end{equation}
such that
\begin{equation}
    \int_{0}^t{H\left(t'\right) dt'} < \mathcal{H} t + N_{\mathrm{max}}
\end{equation}
for all $0 < t < T$. The number of e-folds of expansion is then bounded from above and below by two de Sitter solutions (Fig. \ref{fig:Nvst}),
\begin{equation}
    e^{\mathcal{H} t} < \exp\left[\int_{0}^t{H\left(t'\right) dt'}\right] < e^{\mathcal{H} t + N_{\mathrm{max}}}.
    \label{eq:Nefoldsbound}
\end{equation}
The proper time of a geodesic path on any \textit{past} interval integrating from $t = -t_i$ to $t = 0$ is given by
\begin{equation}
    \Delta s = \int_{-t_i}^{0}{\frac{dt}{\sqrt{1 + v_0^2 a^{-2}\left(t\right)}}},
\end{equation}
with 
\begin{equation}
    a\left(t\right) =  \exp\left[{\int_0^t{H\left(t'\right) dt'}}\right]. 
\end{equation}
From Eq. (\ref{eq:Nefoldsbound}), proper time satisfies a bound $\Delta s_{\mathrm{min}} < \Delta s < \Delta s_{\mathrm{max}}$ (Fig. \ref{fig:svst}), where
\begin{equation}
    \Delta s_{\mathrm{min}} = \int_{-t_i}^{0}{\frac{dt}{\sqrt{1 + v_0^2 e^{-2 \mathcal{H} t}}}}.
\end{equation}
and
\begin{equation}
    \Delta s_{\mathrm{max}} = \int_{-t_i}^{0}{\frac{dt}{\sqrt{1 + v_0^2 e^{-2 N_{\mathrm{max}}}  e^{-2 \mathcal{H} t}}}}.
\end{equation}
From Eq. (\ref{eq:deSitter}), $\Delta s_{\mathrm{min}}$ and $\Delta s_{\mathrm{max}}$ are finite on the interval $-\infty < t < 0$, which means that $\Delta s_{\mathrm{min}} < \Delta s < \Delta s_{\mathrm{max}}$ must also be finite on the same interval. Any spacetime periodic in $H\left(t\right)$ and satisfying the condition $(\ref{eq:EntropyCondition})$ must therefore be geodesically incomplete. 

\section{Conclusions}
\label{sec:Conclusions}

In this paper, we apply the theorem of Borde, Guth, and Vilenkin \cite{Borde:2001nh} to the Ijjas/Steinhardt (IS) model of cyclic cosmology. 
The IS model avoids the problem of entropy growth in cyclic cosmology pointed out by Tolman \cite{Tolman:1931fei} by a Hubble parameter which is periodic in time, but a scale factor which grows exponentially from one cycle to the next. In this paper, we use the BGV theorem to demonstrate that growth in the scale factor inevitably means that the spacetime is geodesically past-incomplete. Therefore any such spacetime is not truly cyclic, in the sense of being perfectly periodic in time. In solving the problem of entropy growth, the model introduces the problem of geodesic incompleteness in its place. This result is completely general: any bouncing spacetime which obeys the condition (\ref{eq:EntropyCondition}) for entropy dissipation and the Null Energy Condition outside the bounce must be geodesically incomplete. This is consistent with the BGV theorem, which shows that any spacetime for which the \textit{average} Hubble parameter is positive must be similarly geodesically incomplete. The IS cosmology satisfies this condition and therefore cannot be past eternal, independent of the details of the dynamics. 

\section*{Acknowledgements}
This work is supported by the National Science Foundation under grant NSF-PHY-2014021.

\bibliography{Paper.bib}

\end{document}